\documentclass[a4paper,11pt]{article}
\usepackage{mathtools}
\usepackage{jheppub} 
\usepackage{graphicx}   
\usepackage{latexsym}   
\usepackage{enumerate}

\newcommand{\SKIP}[1]{}
\newcommand{\fm}{{\rm fm}}                      
\newcommand{\eps}{\varepsilon}			
\newcommand{\bfr}{\mathbf{r}}			
\newcommand{\grad}{\mathbf{\nabla}}		
\newcommand{\EoS}{equation of state}

\title{A machine learning study to identify spinodal clumping in high energy nuclear collisions}

\author{Jan Steinheimer$^1$, Long-Gang Pang$^{2,3}$, Kai Zhou$^1$, Volker Koch$^3$, J{\o}rgen Randrup$^3$ and Horst Stoecker$^{1,4,5}$}

\affiliation{$^1$ Frankfurt Institute for Advanced Studies, Ruth-Moufang-Str. 1, 60438 Frankfurt am Main, Germany}
\affiliation{$^2$ Physics Department, University of California, Berkeley, CA 94720, USA}
\affiliation{$^3$ Nuclear Science Division, Lawrence Berkeley National Laboratory, Berkeley,CA 94720, USA}
\affiliation{$^4$ Institut f$\ddot{u}$r Theoretische Physik, Goethe Universit$\ddot{a}$t Frankfurt, D-60438 Frankfurt am Main, Germany}
\affiliation{$^5$ GSI Helmholtzzentrum f\"ur Schwerionenforschung GmbH, D-64291 Darmstadt, Germany}

\emailAdd{steinheimer@fias.uni-frankfurt.de}

\abstract{
The coordinate and momentum space configurations of the net baryon number in heavy ion collisions that undergo spinodal decomposition, due to a first-order phase transition, are investigated using state-of-the-art machine-learning methods. Coordinate space clumping, which appears in the spinodal decomposition, leaves strong characteristic imprints on the spatial net density distribution in nearly every event which can be detected by modern machine learning techniques. On the other hand, the corresponding features in the  momentum distributions cannot clearly be detected, by the same machine learning methods, in individual events. Only a small subset of events can be systematically differentiated if only the momentum space information is available. This is due to the strong similarity of the two event classes, with and without spinodal decomposition. In such scenarios, conventional event-averaged observables like the baryon number cumulants signal a spinodal non-equilibrium phase transition. Indeed the third-order cumulant, the skewness, does exhibit a peak at the beam energy ($\mathrm{E}_{\mathrm{lab}}= 3-4$ A GeV), where the transient hot and dense system
created in the heavy ion collision
reaches the first-order phase transition.}

\begin{document}
\maketitle

\section{Introduction}
The possible phase transition between a confined chirally broken phase of hadrons and a deconfined phase of quarks and gluons where chiral symmetry is restored has evaded experimental discovery for several decades now. At vanishing net baryon density the transition appears as a smooth crossover, as shown by lattice-QCD simulations \cite{Aoki:2006we,Borsanyi:2010cj,Bazavov:2010sb}. At large baryochemical potentials the situation is less clear as direct calculations on the lattice are not possible due to the sign-problem \cite{deForcrand:2010ys}.
Effective model predictions range from a strong first-order transition to a smooth crossover for the large density domain (or an even more complex phase structure with several transitions) \cite{Stephanov:1998dy,McLerran:2007qj,Alford:2007xm,Motornenko:2019arp}.
In order to determine the phase structure of QCD, several heavy-ion experiments are currently performed or in preparation at RHIC, SPS-CERN, GSI, FAIR, NICA and JPARC-HI.\\

The proposed signals for a phase transition can be roughly split into two categories:
\begin{enumerate}
\item Effects of the softening of the equation of state (EoS) where the appearance of a phase transition leads to a local minimum in the speed of sound. This softening then can be related to changes in the collective flow due to the decreased pressure gradients in the early evolution of the system. Several observables have been suggested ranging from the mean transverse mass to several orders of the azimuthal anisotropies generated in heavy-ion collisions \cite{Hofmann:1976dy,Stoecker:1986ci,Rischke:1995pe,Brachmann:1999xt,Brachmann:1999mp,Csernai:1999nf,Ivanov:2000dr,Stoecker:2004qu,Csernai:2004gk,Nara:2016hbg,Nara:2017qcg,Nara:2018ijw}.
\item Effects from the non-equilibrium features and critical phenomena. In addition to effects of the softening of the equation of state, it is well known that systems can show signals that are related to the appearance of multi-particle correlations. For example, at the critical point the correlation length (in infinite systems) will diverge, leading to characteristic changes in the particle-number fluctuations \cite{Paech:2003fe,Stephanov:2004wx,Stephanov:2008qz,Nahrgang:2011mg,Herold:2013bi,Plumberg:2017tvu,Li:2015pbv}. For systems that undergo a phase transition, formation of baryon clusters can occur due to the spinodal decomposition associated with the mechanically unstable region of the phase diagram \cite{Scavenius:2000bb,Palhares:2010be,Herold:2013qda,Li:2016uvu,Chomaz:2003dz,Randrup:2003mu,Sasaki:2007db,Steinheimer:2012gc}. See also \cite{Sun:2018jhg} for other works on effects of clumping due to a phase transition. 
\end{enumerate}

Recently it was suggested that a new approach based on modern machine-learning methods may offer a new promising venue. As modern neural networks are powerful tools for extracting information from complex datasets, it was suggested to use them to circumvent the biased `handcrafting' of observables. Instead, the neural network should itself select the appropriate features within the data which are most sensitive to the properties of the equation of state. Indeed, foundational work \cite{Pang:2016vdc} has shown that this is feasible, at least in a state-of-the-art relativistic fluid-dynamical approach. The machine learning methods can be employed for the classification of single events, which is different to previous attempts using conventional event-averaged observables. This makes these methods interesting for the analysis of specific features in these individual events.\\

In the present paper we will extend this method to try to identify special phase space features of a first order phase transition, namely features that should appear through instabilities in domains away from phase equilibrium, which are expected to occur in nuclear collisions. In particular we will show that the machine learning methods are able to find these features in the coordinate space but not in momentum space distributions of individual events. The further analysis, using unsupervised learning, supports the idea that these features should be identifiable by event-averaged statistical quantities like the cumulants of the net baryon number distribution.

\section{Method}\label{method}

First we define a framework that is capable of correctly reproducing the underlying physics of the conjectured spinodal decomposition at a QCD phase transition and identifying the appropriate physical observables. In the present work, we employ relativistic fluid dynamics augmented with a gradient term to ensure the proper dispersion relation as expected for spinodal decomposition. In addition we implement an equation of state that is mechanically unstable in the phase-coexistence region at large densities. Such a model has been presented in previous works \cite{Steinheimer:2012gc,Steinheimer:2013gla,Steinheimer:2013xxa,Steinheimer:2017dpb}. Simulations with this model have shown significant baryon clumping due to the spinodal decomposition during the passage of the unstable region in the phase diagram. 

The time evolution of the system
is based on the equations of relativistic ideal fluid dynamics,
namely local four-momentum conservation,
\begin{equation}
\partial_{\mu} T^{\mu \nu} =0\ ,
\end{equation}
and local flavor conservation,
\begin{equation}\label{consr}
\partial_{\mu} j^{\mu}=0\ .
\end{equation}
In this paper we include only the net baryon number current
which is expected to carry most of the spinodal instability strength.
These equations can be solved numerically on a 3+1 dimensional Cartesian lattice \cite{Rischke:1995ir}. In order to take into account the effects of finite-range interactions
(which, for example, are responsible for the presence of a surface tension),
a gradient term is included.
It modifies the equation of state $p(e,n)$ locally,
\begin{equation}\label{a}
p(\bfr)	=p_0(\eps(\bfr),\rho(\bfr))
-a^2{\frac{\eps_s}{\rho_s^2}}\rho(\bfr)\grad^2\rho(\bfr)\ ,
\end{equation}
where $p_0(\eps,\rho)$ is the \EoS \ in equilibrium, i.e.
the pressure in uniform matter characterized by the energy density $\eps$ 
and by the net baryon density $\rho$.
Furthermore, $\rho_s=0.153/\fm^3$ is the nuclear saturation density
and $\eps_s\approx m_N\rho_s$ is the associated ground state energy density.
The strength of the gradient term is conveniently
governed by the length parameter $a=0.033$ fm \cite{Steinheimer:2012gc}. As discussed in detail \cite{Steinheimer:2013gla}, this choice of $a$ results in the formation of
spinodal clusters of a characteristic size of $1-2\,\rm \ fm$.

This model describes nuclear collisions at various incident beam energies. As strong deviations from thermal equilibrium appear in the initial penetrating phase, the ideal fluid dynamical description is supplemented by a non-equilibrium description for the initial state. For this, the non-equilibrium transport model UrQMD, which has been shown to successfully describe a wealth of data \cite{Bass:1998ca,Bleicher:1999xi,Petersen:2008dd} but does not include any effects of a phase transition, is used.

In our calculation the spinodal instabilities are seeded by the fluctuations generated by the initial
UrQMD evolution. In principle there are additional seeds from thermal noise, which, however, are
small corrections to the dominant effects from the UrQMD fluctuations, as discussed in
\cite{Herold:2013bi,Steinheimer:2013xxa}. Since the purpose of the paper is the detectability of
these fluctuations rather than their precise determination, we omit thermal noise in our calculation.

Once the local density of particles reaches a certain value (usually below the coexistence density) the fluid dynamical description looses its validity and the system undergoes freeze-out. At this point we employ the Cooper-Frye procedure, based of the following integration over a predefined hypersurface $\Sigma$ \cite{Cooper:1974mv},
\begin{equation}\label{cooper-frye}
E \frac{dN}{d^3p}=\int_\sigma f(x,p) p^\mu d\Sigma_\mu \, .
\end{equation}

\begin{figure}[t]	
\center 
\includegraphics[width=0.6\textwidth]{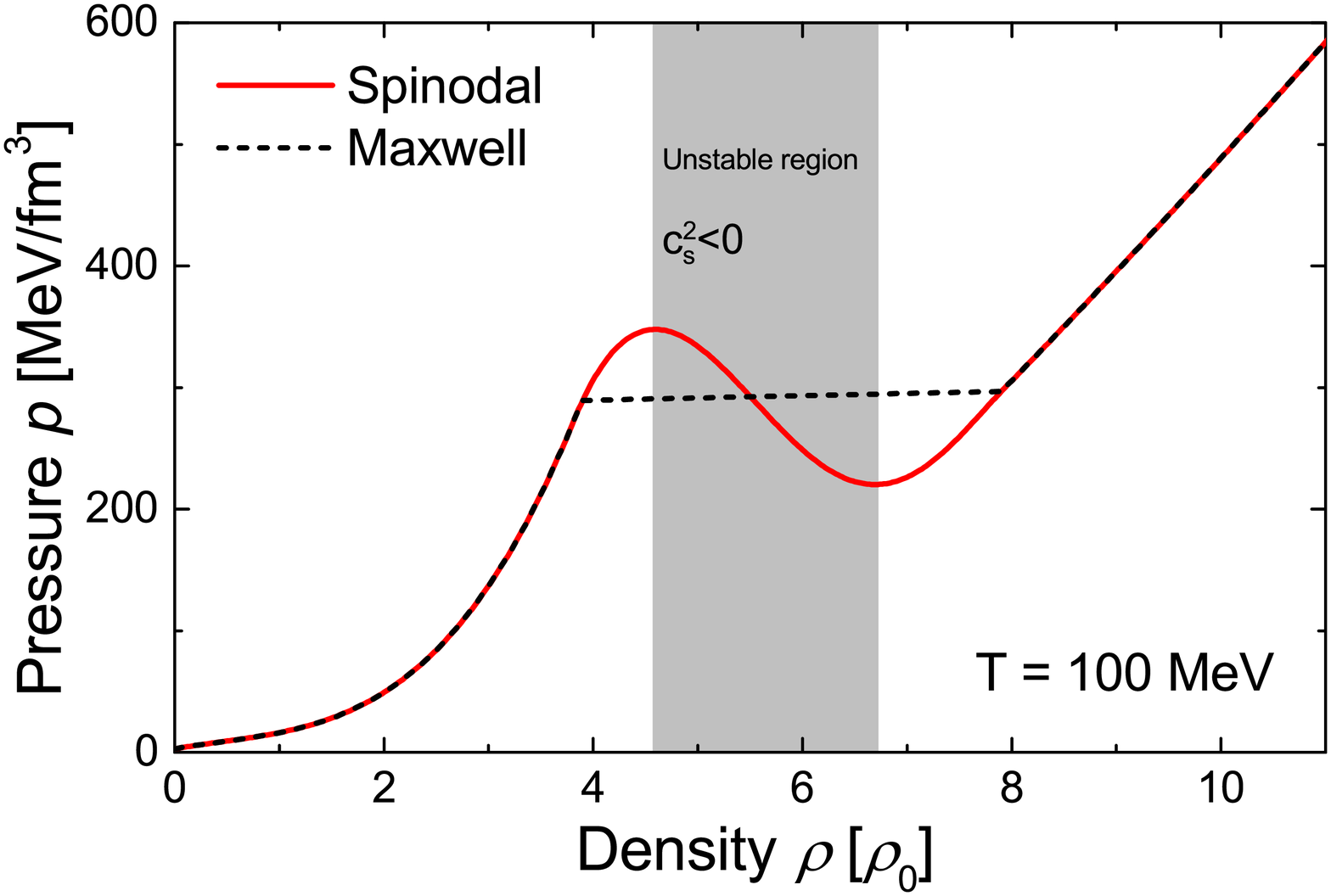}
\caption{\label{f1}Pressure as a function of the net baryon density 
at a fixed temperature $T=100$ MeV,
for both the mechanically unstable spinodal equation of state (solid)
and the corresponding stable equation of state (dashed) where the 
instabilities are removed by means of the standard Maxwell construction.
}
\end{figure}		

At this point, a decision must be made about how to implement the Cooper-Frye 
freeze-out transformation. In most studies, a random sampling of particles on the hypersurface is performed, where the local particle densities, calculated by eq. (\ref{cooper-frye}), are interpreted as probabilities for producing particles. Because the particle numbers per hypersurface element are usually significantly smaller than one, it is assumed that the probability for finding $N$ particles in a cell follows a Poisson distribution, thus ignoring local correlations within a cell. In such a scenario one can enforce global conservation of flavors, which leads to a multinomial probability distribution for finding $N$ particles in a finite number of cells \cite{Huovinen:2012is}.

Alternatively, the Cooper-Frye distribution can be implemented for each hypersurface element exactly. In such an approach, a non-integer particle number will then be emitted from any hypersurface element. Even though this is impractical, it is the only way to conserve all quantum numbers locally exactly (within each single hypersurface cell). In a recent paper it was shown that this local exact conservation reduces significantly the observed fluctuations \cite{Steinheimer:2017dpb}. 

Other methods have also been discussed recently \cite{Oliinychenko:2019zfk}, where the relevant flavors are conserved over clusters of fluid-cells of varying size. 

\begin{figure}[t]	
\setlength{\abovecaptionskip}{-40pt plus 0pt minus 0pt}
\includegraphics[width=0.48\textwidth]{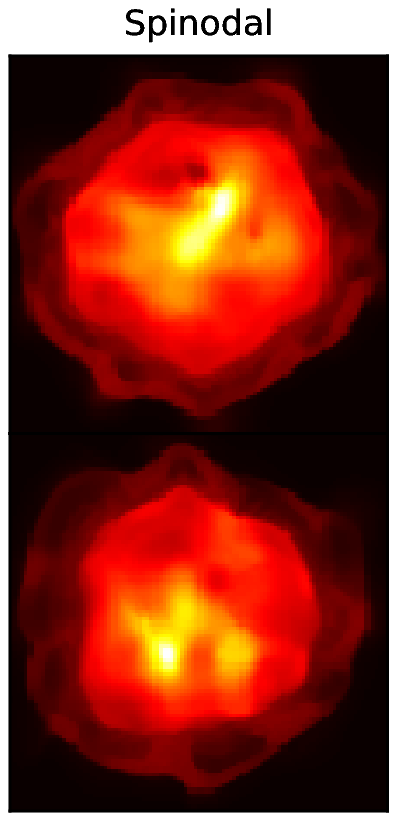}
\includegraphics[width=0.48\textwidth]{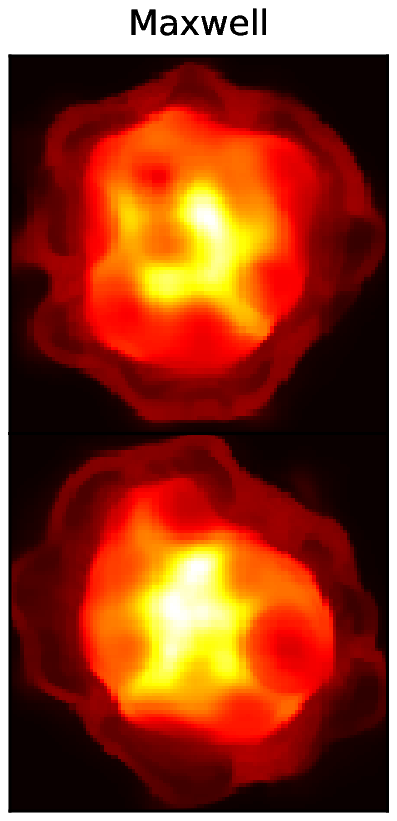}
\caption{\label{bf2}Example images of the normalized transverse density distributions. 20\,000 of these individual Pb+Pb collision events are generated with the spinodal EoS (left) and Maxwell Eos (right). 
These images are used as input to train a convolutional neural network.}
\end{figure}		

It is not yet clear which of these methods should be favored, and how flavors are conserved locally in heavy-ion collisions. As shown in \cite{Steinheimer:2017dpb} the strength of the fluctuations depend on whether at the freeze-out
flavors are conserved locally or globally. In case of one test particle, flavors are conserved only
globally, while with increasing number of test particles flavor conservation becomes more local, as
discussed in \cite{Steinheimer:2017dpb}.  To address both limits of flavor conservation we consider
the freeze-out with one and twenty testparticles. In either case we do {\em not} evolve the system
after freeze-out with hadronic transport.
Thus, we will compare results obtained with a varying number of test particles in the Cooper-Frye procedure. In this way, we will be able to take into account different scenarios of hadron production: 

\begin{enumerate}
    \item One test particle per real particle, {\tt TP=1}: 
    Only global flavor conservation. 
    \item Twenty test particles per real particle, {\tt TP=20}: 
    This scenario comes close to local conservation of flavors, 
    although non-integer particle numbers occur.
\end{enumerate}

In the present work all (test) particles are sampled from the Cooper-Frye equation (\ref{cooper-frye}) and stored in an array. The list of all possible hadrons that can be sampled includes all stable ground-state hadrons, as well as an extensive list of unstable hadronic resonances. After all particles are sampled, the decays of these unstable resonances are calculated within the UrQMD transport model, using the complete Particle Data Group tables as input for the properties of the hadrons and their microscopic decays. Each test-particle resonance then decays exactly as a regular resonance, where the decay products carry only a fractional flavor.

With this model, heavy-ion collisions can be modeled at any beam energy which provides sufficient compression and heating to allow for a coexistence of confined and deconfined matter. To find out where the conditions for observable signals is best, the beam energy can be varied.

In the following section, we discuss the equation of state, 
the most important physics ingredient in the present simulations.

\section{The equations of state}

As we seek to identify signals for the predicted spinodal decomposition, or baryon clumping, we will focus on two equations of state that differ only with respect to the
instabilities associated with the phase transition.
Hence, they are identical outside of the spinodal region of the phase diagram, 
but within that phase coexistence region they differ significantly.

The spinodal EoS has a mechanically unstable region 
with a negative square of the isothermal speed of sound $c_s^2<0$.
The stable partner EoS is obtained by means of a Maxwell construction
(which has no effect in the already stable phase regions).
The two equations of state are illustrated in Fig.\ \ref{f1}
and more details on the construction of these equations of state
may be found in Ref.\ \cite{Randrup:2009gp}.
Well within the confined and deconfined phase regions,
these equations of state describe a gas of interacting nucleons and pions
and a gas of free two-flavor quarks and gluons in a bag, respectively.

Previous work has shown that the spinodal EoS and its Maxwell
partner EoS lead to similar collective radial expansions 
\cite{Steinheimer:2013gla}. 
By virtue of its construction,
the Maxwell EoS produces the same amount of work during the expansion, which is proportional to $\int p dV $, as does the spinodal EoS, hence the amount of energy transformed to collective motion is exactly the same in both EoS cases.

In principle, there are many model predictions for the high baryon density equilibrium equation of state and here
we adopt that of Ref. \cite{Steinheimer:2012gc}. As the present paper focuses on the detectability of the spinodal
clumping associated with a first order phase transition, any equation of state that
exhibits a co-existence between hadronic and quark matter and its associated spinodal region would
serve this purpose.

We expect that any dynamical differences between the two scenarios are
to be associated to the coordinate space correlations,
in particular to the degree of baryon clumping during the phase separation. 
Indeed previous studies have shown that the two equations of state yield statistically significant differences in the baryon number distributions \cite{Steinheimer:2017dpb}.
On the other hand, 
it has not been established whether the spinodal decomposition mechanism
leads to clumping in essentially every event or in only a few.
The main unanswered question is whether the clumping in coordinate space will actually yield a measurable significant signal in momentum space.

\section{Using Deep Learning}

In this paper, several popular machine- and deep-learning methods are 
applied in order to determine whether it is possible to discriminate
between those events that are generated by the fluid dynamical model through the non-equilibrium spinodal EoS and those events that are generated by the Maxwell EoS corresponding to an equilibrium phase transition.
To accomplish this goal, we first compare two different neural network architectures, which represent different supervised learning approaches.
The first one is a convolution neural network (CNN), 
where event-by-event images of the density distribution 
in coordinate as well as momentum space serve as input. 
CNNs are used successfully in pattern recognition tasks in image applications.
The second model is a point cloud network (PCN) \cite{qi2016pointnet}, 
whose inputs are lists of discrete particle properties, 
e.g.\ particle four-momenta for every individual particle in a single event.
The PCN is well suited for dealing with particles 
from collision experiments because it can use the momentum information 
for discrete particles as direct input. For a short introduction on neural networks and terminology we refer to appendix \ref{intro_nn}.

\begin{figure}[t]	
\center \includegraphics[width=0.6\textwidth]{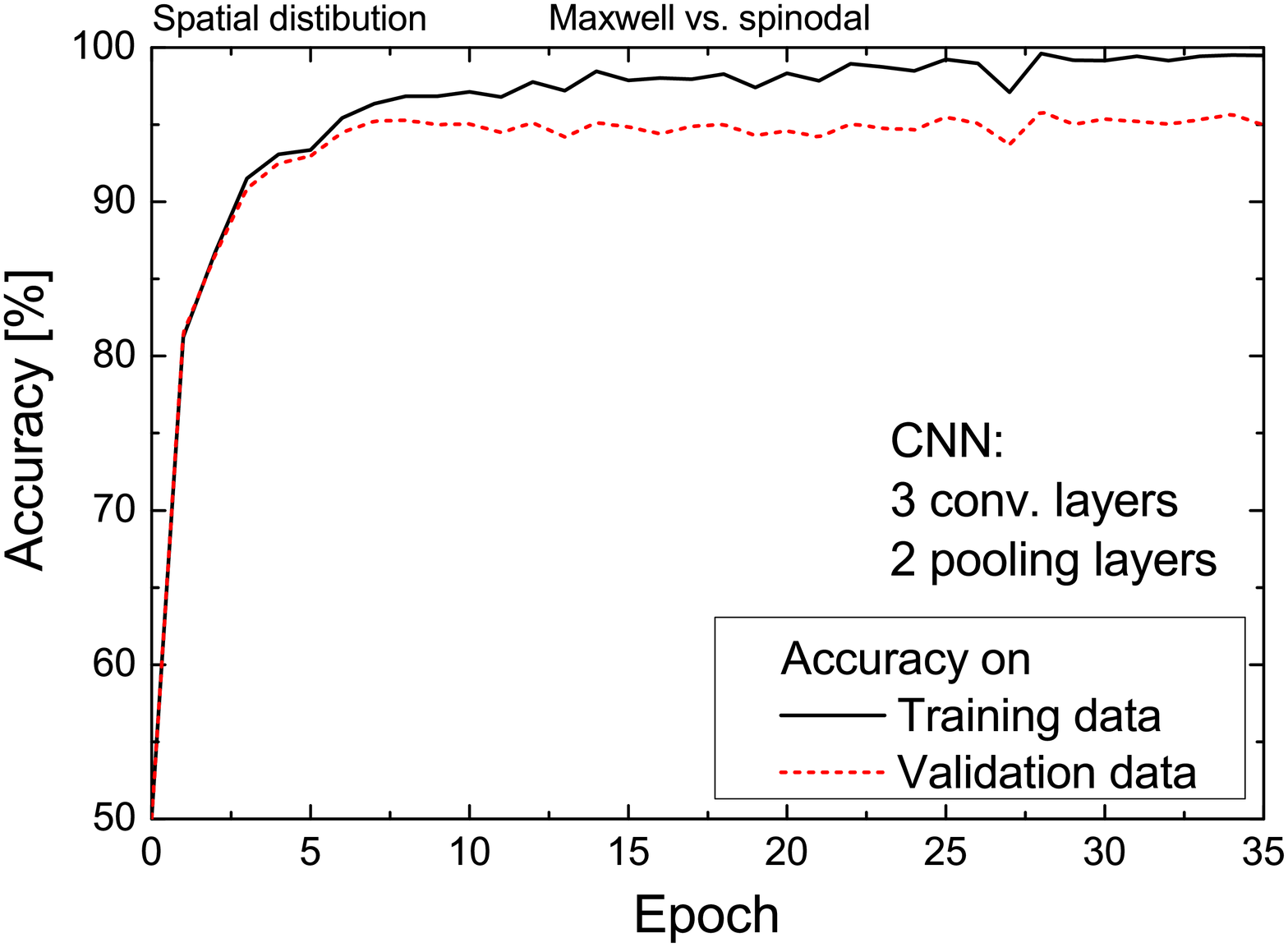}
\caption{\label{bf4}
Training and validation accuracy for 
the Convolutional Neural Network (CNN) with coordinate-space input. 
After about 5 epochs the network starts over-fitting. 
Nevertheless, a good accuracy of about $95\%$ can be achieved 
without optimization of the network structure.}
\end{figure}		

The last section presents a unsupervised learning approach, i.e. a principal component analysis (PCA), 
which is used to extract the principal components of a given analysis feature, namely the two-particle momentum difference distributions (see appendix \ref{pca_intro} for details on the PCA). 
This feature is fed to a fully connected neural network (NN) to identify the EoS. The PCA yields a slightly improved accuracy.

\subsection{Coordinate space}
\label{s_coord}

In a first step, we test the neural network for the coordinate clumping as expected from the spinodal equation of state: About 20\,000 Pb+Pb collision events are generated at a (typical FAIR/GSI) beam energy of $E_{\mathrm{lab}}=3.5\, A$~GeV, for each EoS. 
We know from previous studies of the moments of the density distribution \cite{Steinheimer:2012gc} that the density fluctuations in
coordinate space are strongest at $t=$3~fm/$c$ \footnote{The precise value of the optimal energy, $E_{\rm lab}=3.5\rm \,GeV $, depends somewhat
on the specific equation of state employed, so others would yield different values.}
at this beam energy and subside after another 3~fm/$c$. Thus
we stop the time evolution of the system at the point in time where the density fluctuations are expected to be strongest, at $t=$3~fm/$c$. 
From each event an 'image' is then generated, containing
information on the net baryon density distribution 
in the transverse spatial $X-Y$ plane for $Z=0$. 
A naive way to classify the two scenarios is to just compare
 the maximum density for all the images and assume/postulate
that the largest densities can be reached only in the spinodal case. 
However, to avoid such a trivial comparison and to make the task more challenging for the neural network, we renormalized the event-by-event density distributions 
by their maximum value for each event separately. 
Examples of the resulting, single event, distributions are shown in Figs.\ \ref{bf2}. 
Each image has a dimension of $100\times100$ pixels 
which corresponds to the number of fluid-cells shown in these figures.

For the training stage, a CNN with three convolutional layers and two pooling layers in between is used (See appendix \ref{cnn_coo} for more details on the network structure).
The output of this network is a binary classification on 
whether an input picture is from a spinodal or from a Maxwell event. 
Figure \ref{bf4} shows the resulting accuracy during the training stage for the training dataset and for an independent validation set. 
It is obvious that the training as well as the testing sets accuracy increase quickly as the network learns the most important features for both cases. 
At later times the network overfits the data, 
indicated by the ever increasing training accuracy. 
In any case, already for the simple network, a $95 \%$ accuracy is obtained for the classification of events. 
This important finding has a consequence: it suggests that the spinodal EoS creates characteristic features in almost every event in coordinate space, which can be discriminated from features in the Maxwell case.

We speculate that these features correspond to the actual density clumping, however the network does not reveal how it reaches its results.
Given the underlying physics, our speculation seems to be a reasonable explanation.

Note that this finding confirms previous findings. Indeed
 spinodal instabilities lead to characteristic structures 
in coordinate space \cite{Steinheimer:2017dpb,Paech:2005cx}. 
It is most important to point out that we have now verified that these clumping structures appear in nearly every sampled event.

\subsection{Momentum space}\label{s_mom}

\begin{figure}[t]	
\center\includegraphics[width=0.6\textwidth]{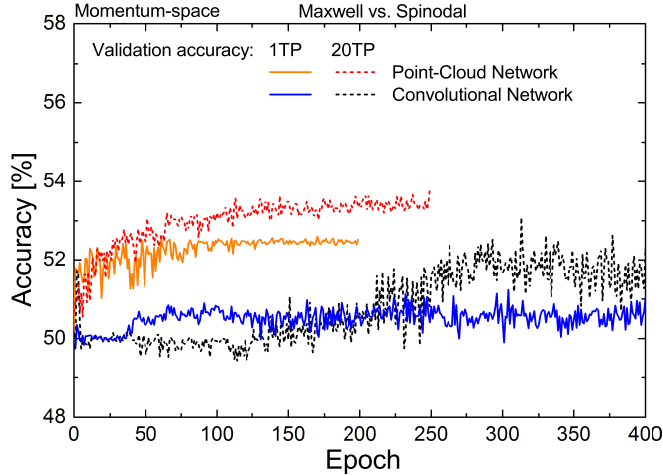}
\caption{\label{bf5}
Validation accuracy for the CNN and the PCN,
comparing the spinodal EoS and the Maxwell construction 
for {\tt TP=1} and {\tt TP=20}.
The PCN shows a better performance than the CNN,
but both network structures display only a very low validation accuracy 
just above $50\%$.}
\end{figure}		

Next let us focus on the event-by-event baryon distributions in momentum space. Here the connection of the final momentum space distributions to the baryon clumping during the early compression phase evolution is less obvious. In order to obtain the 'final' information on the momenta of all produced particles we need to run the fluid dynamical simulation until a later time. Although one may naively argue that the momentum space correlations should be largest at the point in time when the coordinate space clumping is large, this is actually not the case. In order to transform the coordinate clusters to momentum space correlations, these clusters need to expand fluid dynamically. Furthermore, the dense clusters are composed of dense quark matter which cannot be detected by experiment.
In a realistic scenario, baryons will in fact be produced on an isoenergy density hyper-surface that is below the coexistence density at $e=4 e_0$ with $e_0$ being the nuclear saturation energy density.
On this isoenergy hypersurface, the baryon density is approximately constant. and thus the signal in coordinate space has vanished completely. It is however possible that the coordinate space clumping has been transformed to momentum space correlations by the fluid dynamical evolution. Consequently, at the point of particle production, the clumping in coordinate space has disappeared and we have to rely on correlations in momentum space 
to discriminate between the two event classes. 

The baryons are produced by a random sampling of the Cooper-Frye hypersurface as discussed in section \ref{method}, 
either sampling the actual number of particles ({\tt TP=1}) 
or a large number of test particles ({\tt TP=20}) per real particle. 
For both cases, all resonance decays are performed and the resulting phase space information for all produced particle (real and test particle) is saved in a list.\\
This list can be used to create a two-dimensional histogram of all baryons, using equal-sized bins in $p_x$ and $p_y$. The histograms dimension are 20$\times$20 bins of width 200 MeV in the range of $-2<p_{x,y}<2$ GeV. This histogram then gives an image, similar to the coordinate space image, but more coarse grained due to the finite number of particles. For the case of {\tt TP=20}, 120\,000 training events and 20\,000 test events were generated.

Again a CNN (see Appendix \ref{cnn_mom} for the network structure)
was trained to classify the equation of state on an event-by-event basis. The results on the training and testing accuracy are shown in figure \ref{bf5} as black and blue lines. It is clear that the network fails to find significant features that would allow it to discern the spinodal from the Maxwell class on an event-by-event basis. While for {\tt TP=20} an accuracy of $52\%$ can be reached,
 the accuracy for {\tt TP=1} is almost as low as $50\%$ 
which would mean that the network result is no better than random guessing.
One may argue that the 20$\times$20 bin size input is to coarse grained to capture the relevant correlations, however, increasing the number of bins does not lead to a better result as it significantly increases the noise on the event-wise input data.

\begin{figure}[t]	
\center \includegraphics[width=0.6\textwidth]{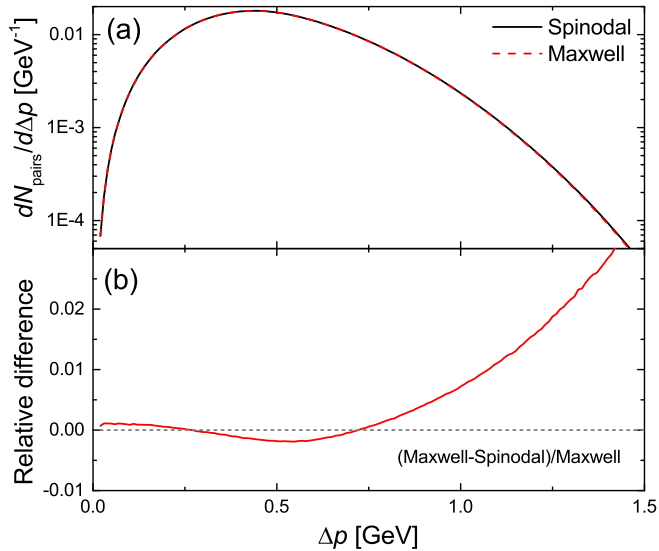}
\caption{\label{f_dnddp} Panel $a$:
The distribution of the momentum difference between baryon pairs 
after the Cooper-Frye hadronization procedure;
Both, the spinodal and Maxwell distributions, are very similar.
Panel $b$: The small relative difference between the distributions.}
\end{figure}		

The CNN is a state-of-the-art pattern recognition method when the input is an image-like instance.
In this particular case, the input 'images' are created from the momentum density distribution of baryons.
More precise momentum information about the baryons could be lost during the data processing and one would prefer to actually use all particle information as training input.

To use the full information about all the discrete particles, 
we also employ a point cloud network architecture taking as input points in a five-dimensional feature space (for more details on the network structure and the hyper-parameters used, see Appendix \ref{pcn}).
The five features $(E, p_x, p_y, p_z, {\rm mass})$ 
define one point for each particle. 
Thus the PCN is more convenient and better suited for dealing with lists of particle information as it is obtained from experiments and event generators.
An important advantage of this network is that 
the particle lists of all events have a permutation symmetry, 
i.e. changing the order of particles in an event does not affect the final result.

With the PCN, the testing accuracy can now reach $52\%$ for the {\tt TP=1} case 
and $54\%$ for {\tt TP=20}, as shown in Fig.\ \ref{bf5}. 
Even though the PCN is more convenient for dealing with lists of particles 
and does an overall slightly better job at the event classification, the accuracy 
is still rather poor when only momentum-space information is taken into account. 

\subsection{Using different features}\label{mom_diff_s}

\begin{figure}[t]	
\center\includegraphics[width=0.6\textwidth]{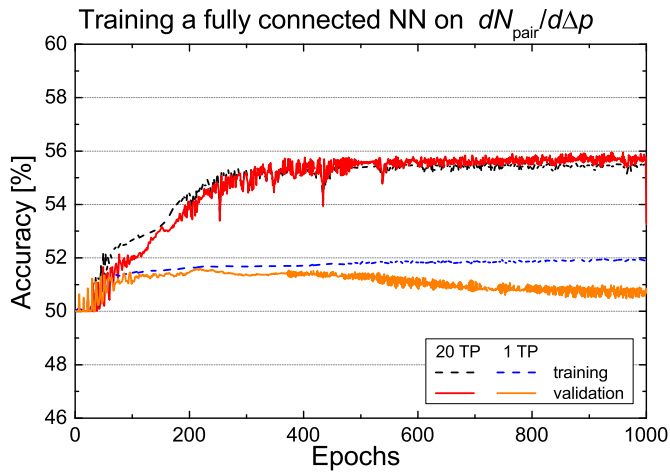}
\caption{\label{f_ann}
Training and validation accuracy as function of the number of training epochs for the fully connected neural network, training with event-by-event distributions of the momentum difference of baryon pairs. We compare results for {\tt TP=1} and {\tt TP=20}.
}
\end{figure}		

An important result of the last section is that the CNN as well as the PCN are not able to efficiently extract distinguishing features from the momentum space information, i.e. the plain $p_x$-$p_y$ spectrum or the baryon momentum vectors. As a result the classification accuracy is very low, only 52$\%$.

Since the underlying hope behind the consideration of spinodal clumping is that the clusters will produce correlations in momentum space, it may be useful to construct a distribution of momentum differences instead of the pure spectrum. 
This feature engineering may yield better results than leaving the full engineering to the neural network. 

The two-particle momentum-difference distribution $dN_{\mathrm{pairs}}/d\Delta p$
can be calculated event-by-event by binning the momentum difference
between all pairs of baryons $a$ and $b$ in the event,
\begin{eqnarray}
\Delta p_{ab} &\equiv& \mbox{$\frac{1}{2}$}|\mathbf{p}^a-\mathbf{p}^b|\\ \nonumber
&=& \mbox{$\frac{1}{2}$}\left[(p^a_x -p^b_x)^2+(p^a_y -p^b_y)^2+(p^a_z -p^b_z)^2\right]^{\frac{1}{2}} .
\end{eqnarray}

The quantity $\Delta p_{ab}$ is equal to the momentum of each of the two particles
as measured in the rest frame of the pair.
An earlier study \cite{RandrupHIP22} considered a somewhat similar observable,
namely the average kinetic energy of each particle in an $N$-body cluster
and found that the signal-to-background grows stronger as the cluster size $N$
is increased but, at the same time, the counting rate decreases progressively.
In the present exploratory study, we stick to just two-baryon clusters.

The resulting event-averaged distributions for the spinodal and Maxwell cases are shown in the upper panel of Fig.\ \ref{f_dnddp}, while the relative difference of these distributions is shown in the lower panel of Fig.\ \ref{f_dnddp}. It should be noted that this distribution gives identical results for different numbers of test particles, but the amount of noise is larger when the number of test particles is small. The relative difference of these baryon pair distributions is small, but appears systematic.
In the spinodal case, intermediate-momentum differences ($\Delta p< 0.5$ GeV) are preferred, while large-momentum differences are suppressed. In order to find out whether these small differences can be used to distinguish the event classes, we will employ a fully connected neural network for classification. This network structure is chosen due to the simpler input data, namely the $dN/d\Delta p$ distribution,
which is a 200-bin dataset.
After training both the {\tt TP=1} and {\tt TP=20} datasets, we find that indeed the neural network performs better than on the pure $\left\{p_x, p_y\right\}$ spectra. However only an accuracy of 55$\%$ for {\tt TP=20} and 52$\%$ for {\tt TP=1} is reached. Even though this is better than random, it is not sufficient to claim the discovery of a two class distribution.

\begin{figure}[t]	
\center\includegraphics[width=0.6\textwidth]{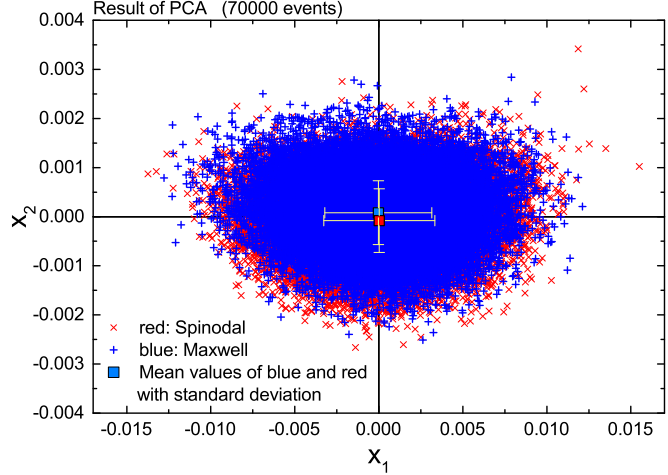}
\caption{\label{f_pca}
Scatter plot of the first two components of a PCA 
of the distribution of the momentum difference for baryon pairs. 
The red crosses indicate the events with the spinodal instabilities, which dominate the southern hemisphere, 
while the blue pluses indicate events with a Maxwell construction. 
The large symbols with error bars near the center indicate the mean values with their standard deviation. The spinodal EoS creates a clear crescent of crosses in the southern hemisphere, $x_2<0$. Also the mean value is shifted downwards slightly. }
\end{figure}		

\begin{figure}[t]	
\center \includegraphics[width=0.6\textwidth]{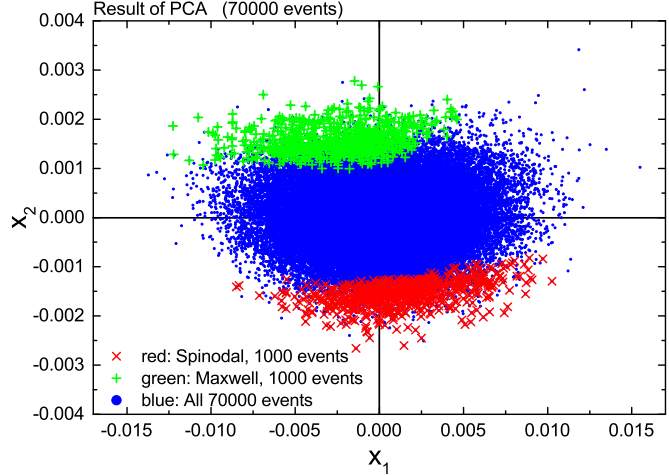}
\caption{\label{f_pca_2}
Scatter plot of the first two components of a PCA 
of the distributions of the momentum difference for baryon pairs. 
The blue points correspond to all 70\,000 events, Maxwell and spinodal. 
The red crosses correspond to those events that were identified correctly
as being in the spinodal class from among those 1000 events 
that had the highest probability for being spinodal events,
according to the neural network.
Similarly, the green pluses show the correctly identified Maxwell events
among the 1000 event having largest probability of belonging to that class.
According to this distribution it is reasonable to assume that the neural network would assign points (distributions) in the norther hemisphere a larger probability to be from the Maxwell EoS and points in the southern hemisphere to be spinodal events. Therefore it is clear that the $x_2$ variable serves as discriminator for the 
two event classes and that the network result is dominated by the small shift in an $x_2$-like feature.}
\end{figure}		

\subsection{Unsupervised learning}

In order to better understand why only a relatively poor individual event 
discrimination can be obtained, we will analyze the $dN/d\Delta p$ distribution using a simple unsupervised learning tool, because the distribution is only one dimensional. Here we will use a principal component analysis (PCA) which extracts the most relevant features of a function.
To understand the results, one should keep in mind that for a machine-learning analysis the  $dN/d\Delta p$ distribution is a 200-dimensional vector, i.e.
 each $\Delta p$ bin corresponds to a dimension in the training feature vector. The PCA analysis will try to reduce the number of dimensions of the feature vector by transforming the basis of the vector such that the variance in the first components of the vector (for example a two-dimensional vector) is maximized. It is trying to preserve the maximum amount of information. Such principal components might be related to the peak position or width of the distribution or some combination of these.

Figure \ref{f_pca} shows a plot of the first two principal components $x_1$ and $x_2$ for all training events with {\tt TP=20}. Each point corresponds to a single event and the color indicates whether the event corresponds to the spinodal EoS (red crosses) or the Maxwell construction (blue pluses). Note that in the PCA this information was not given. The large square symbols denote the average values of the components of a specific event class. In addition, the standard deviation of the extracted components are shown as error bars.   

The figure indicates that the mean values are similar with only a small systematic shift. It was checked that this shift is statistically significant (the error of the mean is much smaller than the difference of the means).
A closer examination reveals that the red crosses (spinodal) are more frequent for negative values of $x_2$. This indicates that the spinodal EoS more likely produces events that `look' different than the average event. On the other hand, these events are rare which explains why an individual event classification analysis did not produce the desired results. In other words, only few events display characteristic features, while most events appear indistinguishable.

\begin{figure}[t]	
\center \includegraphics[width=0.6\textwidth]{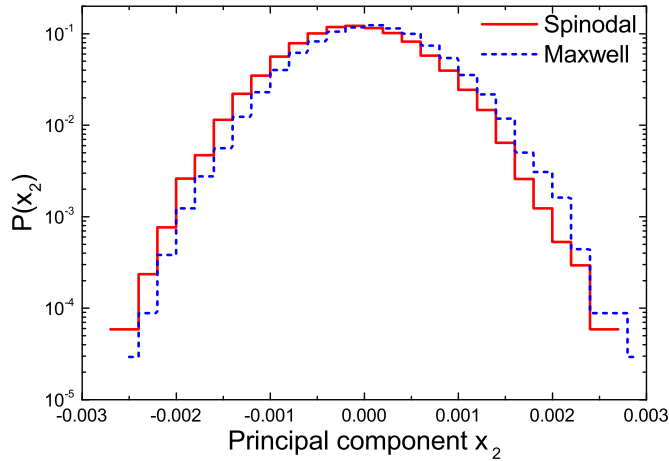}
\caption{\label{x2_d} Distribution functions of the second principal component $x_2$ for both classes, \textit{Maxwell} (blue dashed) and \textit{spinodal} (red solid). One can clearly see the small shift in the mean $x_2$ and the effect on the tails of the distributions. }
\end{figure}		

\begin{figure}[t]	
\center \includegraphics[width=0.6\textwidth]{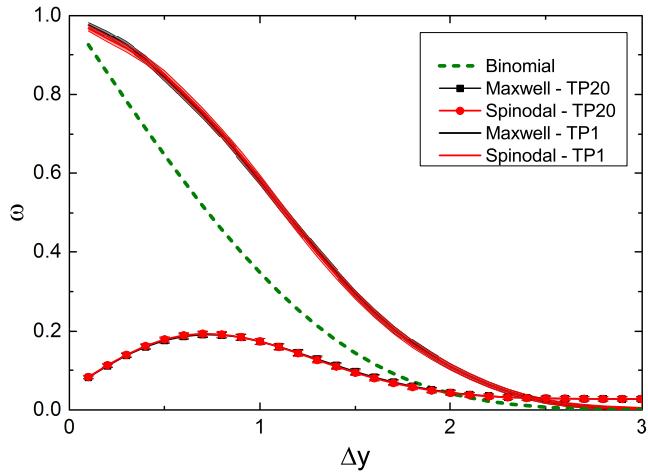}
\caption{\label{af1}
The scaled variance of the net baryon number as a function of the rapidity window size around mid-rapidity. Different scenarios are shown: 
The results for {\tt TP=1} and {\tt TP=20} with spinodal and Maxwell EoS, 
as well as a binomial baseline.}
\end{figure}		

Fortunately there is a way to verify this suspicion. The output of the neural network, before the final \textit{Softmax} function, are actual values that are proportional to the probability that a given input event belongs to either class Maxwell or spinodal. Thus the network assigns each input a probability that it belongs to either class, that is then transformed into a binary decision. By studying how these probabilities are assigned, with the help of the unsupervised learning approach, we can learn about the networks decision process. First, we selected those 2000 events to which the fully connected neural network had assigned the largest probabilities for belonging to either class, spinodal or Maxwell. For these events we plotted in Fig.\ \ref{f_pca_2} only those that are correct classifications as spinodal (red crosses) and Maxwell (green pluses). The blue points refer to the total dataset. It is obvious that the neural network separates the events mostly according to the $x_2$ feature also found in the PCA. It turns out that the accuracy for those events shown (the 2000 events with the largest probabilities) is around 70$\%$. \footnote{Again, accuracy refers to the likelihood that the network makes the correct decision, while the probability refers to how likely the network thinks an event belongs to either class.}
This is considerably better than for the total dataset, which was around 54$\%$. This finding confirms the intuitive suspicion that there is a strong overlap in the features of the spinodal and Maxwell events. The network thus focuses on events that are outliers, having the strongest features, 
namely $x_2$ in our case.
Consequently, only a relatively poor accuracy for all events can be obtained. 

We note that the fact that the network has the highest accuracy for the events in the tails of the
  $x_{2}$ distribution of the PCA (green and red points of Fig. \ref{f_pca_2}  does not mean that the
physics for these events is different than that for those in the same event class with a smaller value of
$x_{2}$. As shown in Section 4.1 almost all events with the spinodal EOS show characteristic features in coordinate
space. However, the correlations, which are clearly visible in coordinate space, are rather weak in
momentum space, so that the $\Delta p$ distribution is dominated by noise. This is illustrated in
Fig. \ref{x2_d}, where we show the $x_{2}$ distribution for both event classes. While the peaks are
clearly separated their widths are very large so that only the events in the respective tails can be
uniquely associated with a given event class.
 
\begin{figure}[t]	
\center \includegraphics[width=0.6\textwidth]{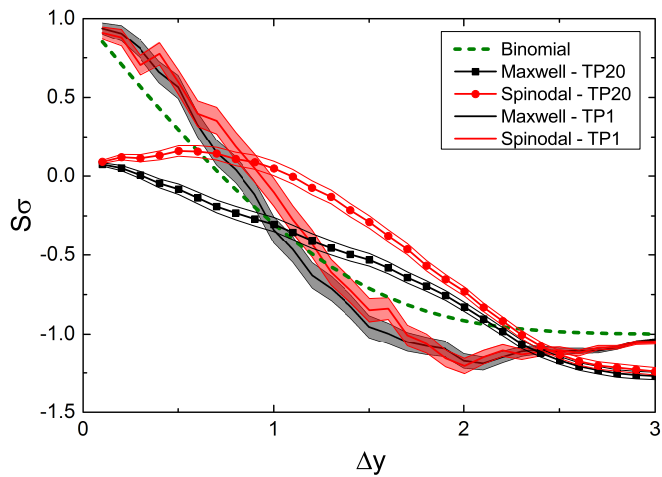}
\caption{\label{af2}The normalized skewness of the net baryon number as function of the rapidity window size around mid-rapidity. 
Different scenarios are shown: The results for {\tt TP=1} and {\tt TP=20}
with spinodal and Maxwell EoS, as well as a binomial baseline.}
\end{figure}		
\section{Conventional Observables} 

In this final part we use `conventional' statistical methods to distinguish spinodal events from Maxwell events. In particular, we construct the cumulants of the baryon number distribution function, as they are also measured in several heavy ion experiments. The advantage of these cumulants is that they are sensitive also to the outlying events.
The first three cumulants are given by
\begin{eqnarray}
K_1 &=& M  = \left\langle {N} \right\rangle\ , \\
K_2 &=& \sigma^2 = \left\langle (\delta {N})^2 \right\rangle\ , \\
K_3 &=& S \sigma^{3} =\left\langle (\delta {N})^3 \right\rangle\ ,
\end{eqnarray}
where $\delta{N}= N -\left< N \right>$, $N$ is the number of particles in a given experimental acceptance window and the brackets denote an event average. These cumulants, and ratios thereof, have been measured by the STAR, ALICE, NA61 and HADES experiments \cite{Aggarwal:2010wy,Adamczyk:2014fia,Arslandok:2015uno}.
Usually, convenient ratios of the cumulants are presented:
\begin{eqnarray}
\omega &=& K_2/K_1\ , \\
S \sigma &=& K_3/K_2 \ .
\end{eqnarray}

In the following we show results from our model simulations for the same events that where used in the machine-learning analysis. We have calculated the net baryon number cumulants in intervals in rapidity, around mid-rapidity, integrating over the entire transverse momentum distribution. The errors are estimated using the delta-theorem and a random distribution, which should give an upper estimate of the error (for more details on the error see e.g. \cite{Luo:2017faz,Bzdak:2018axe}).

Figure \ref{af1} shows the scaled variance $\omega$
of the baryon number as a function of the rapidity window size for the two scenarios with {\tt TP=1} and {\tt TP=20}. A clear difference between the two cases is observed, but the scaled variance for the two cases, spinodal and Maxwell are almost identical. 

Figure \ref{af2} shows $K_3/K_2$, 
for the same rapidity window size.
Again the results show a clear dependence on the number of test particles used,
whether twenty ({\tt TP=20}) or one ({\tt TP=1}).
Importantly, for this observable the two EoS scenarios considered
lead to distinguishable results,
the difference being quite significant for {\tt TP=20}
and still visible for {\tt TP=1}.

\begin{figure}[t]	
\center \includegraphics[width=0.6\textwidth]{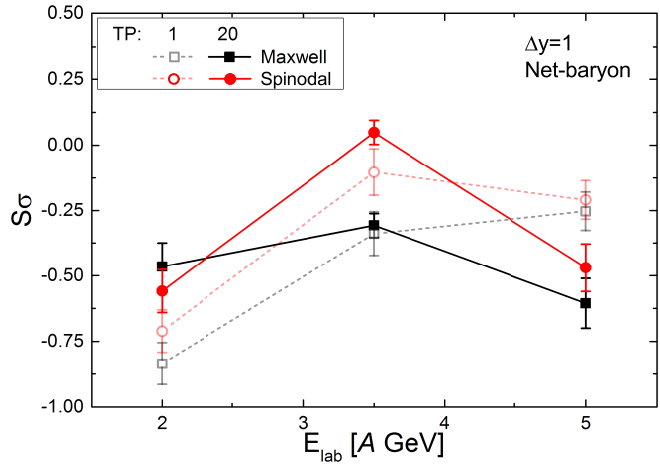}
\caption{\label{af3}
The normalized skewness of the net baryon number distribution in the rapidity window of $-0.5<y<0.5$ for several incident beam energies (in the lab frame). 
Results for {\tt TP=1} and {\tt TP=20} are compared. 
A peak is found for the beam energy that produces 
the largest effect of the spinodal decomposition. 
}
\end{figure}

Finally, the beam energy dependence of the skewness is presented in Fig.\ \ref{af3}. A clear enhancement is seen in this observable at the beam energy with the strongest clustering ($E_{\mathrm{lab}}= 3.5$ A GeV). The peak is strongest for {\tt TP=20},
but the enhancement is still visible for {\tt TP=1}.
We propose that this peak presents an observable indication of
spinodal decomposition in nuclear collisions
due to the occurrence of a first-order phase transition.

\section{Discussion}

We have presented a detailed investigation of how phase-space clumping of the baryon number, due to a first order phase transition, can or cannot be observed in heavy ion collisions. Employing state-of-the-art machine learning methods we were able to show that the QCD phase transition will lead to systematic features, in the coordinate space baryon density distribution, in essentially every event, given that the beam energy is tuned so that the created system will lie in the unstable region of the phase diagram. It was also demonstrated that the translation from coordinate space clumping to momentum space clumping is far from trivial. Using Cooper-Frye sampling of particles as well as test particles, while globally conserving the baryon number, the systematic  distinguishable features of the spinodal clumping, in individual events, almost entirely disappeared in momentum space (which is the only kind of data available to heavy ion experiments).

Both the CNN and the PCN where able to reach only slightly better than random accuracy, discriminating between single spinodal and Maxwell events, on the basis of momentum space correlations. 
Similarly, no measurable discrimination was obtained on the basis of individual event net baryon rapidity distributions.
Furthermore, when the two-baryon correlation function was explicitly calculated and used as training input for a fully connected neural network, small effects on an individual events were found. The accuracy of this method was similar to that of the PCN. The deep learning methods presented in this work where not able to reliably identify the effects of spinodal decomposition when given single event momentum distributions as input. Clearly, it seems very difficult to identify these effects in the momentum distributions of individual events.

It is important to note that we did indeed find a systematic shift of the event properties from Maxwell to spinodal events when a PCA analysis was performed.
We have found that the PCA is most sensitive to the momentum difference distribution
around 500-600 MeV. At the same momentum difference also the average distributions show the largest absolute difference, as seen in figure \ref{f_dnddp}(b). Therefore it is reasonable to assume that the difference in physics lies in a small shift of the
mean of the two proton momentum difference distributions, and those events that have the largest difference in the $\Delta p_T$ distribution around 500 MeV are most likely to be correctly identified.

Consequently, observables that are sensitive to the tails of the event distributions, are more promising than individual event observables. Such observables are for example higher-order cumulants of the net-baryon distribution.

We predict that the transition through the QCD phase transition is visible as a maximum of the skewness of the net-baryon number multiplicity distribution in a central rapidity window of $-0.5<y<0.5$, which is accessible with current and future experiments.

We also expect that higher-order cumulants, such as the kurtosis, should also show effects of the spinodal decomposition, but within the current calculational statistics we could not make reliable predictions. Thus, creating a fluid dynamical model that can incorporate the effects of spinodal decomposition, but has a significantly reduced computational time would be an important future task.\\

\section{Acknowledgments}
JS and KZ thank the Samson AG and the BMBF through the ErUM-Data project for funding. JS and HS thank the Walter Greiner Gesellschaft zur F\"{o}rderung der physikalischen Grundlagenforschung e.V. for her support. This work was supported by the DAAD through a PPP exchange grant.
This material is based upon work supported by the U.S. Department of Energy, Office of Science, Office of Nuclear Physics, under contract number DE-AC02-05CH11231. L.G. Pang is supported by NSF under Grant No. ACI-1550228 within the JETSCAPE Collaboration.
Computational resources where provided by the NVIDIA Corporation with the donation of two NVIDIA TITAN Xp GPUs and the Frankfurt Center for Scientific Computing (Goethe-HLR).

\appendix

\section{Short introduction to neural networks and the training process}\label{intro_nn}
In simple terms one can understand a neural network as some kind of mapping
  function, which maps an $n$-dimensional input vector to an $m$-dimensional output vector. In our
  specific case the output vector is two dimensional, the value of each dimension determining whether the input vector belongs to class 1 (Spinodal) or class 2 (Maxwell). Besides these so-called input and output layers a neural networks consists of a varying number of hidden layers. These hidden layers consist of neurons who themselves perform a non-linear transformation on their input. This non-linear transformation usually consists of a linear transformation $y=a x +b$, where $x$ is the input of the neuron, $y$ is the output of the neuron and $a$ and $b$ are parameters of that specific neuron (in the case of many neurons and many layers $a$ and $b$ take the form of multi dimensional matrices). The output $y$ then serves as argument of a so called \textit{activation function} which can take on different forms (often sigmoid or Relu functions are used). Depending on the structure of the network the neurons of one single layer can be connected to any number of neurons in the next layer. For example in a fully connected neural network, the output of the $j^{\mathrm{th}}$ neuron in the $i+1^{\mathrm{th}}$ layer is the sum of all outputs of the $i$th layer $y_{i+1,j}=f(\sum_{k} a_{k,i+1} x_k +b)$ where $k$ is the number of neurons in the $i^{\mathrm{th}}$ layer and $f()$ is an activation function. As one can see such a network can easily have a large number of parameters to be determined. The determination of these parameters is done during the training phase of the network. During this training, a number of pre-labeled data-points is fed to the network. Using an error-function which measures how well the network output agrees with the true training label (in our case the \textit{cross entropy}), the parameter values are changed using a gradient descent method in order to minimize the error with respect to the training data. Since the gradient descent method only changes the parameters incrementally, many passes through the whole training data sets are necessary in order for the gradient descent to converge. Such a single pass through the training data-set is called \textit{epoch}. Once the network is trained it can be validated using an independent set of data.\\
Due to the large number of parameters a neural network is prone to \textit{overfit} the data,
i.e. the error on the training data decreases as the error on the validation data increases. A
popular regularization method to avoid overfitting is the \textit{dropout}.\\
For the \textit{dropout} an additional layer of $i$ neurons is added between the hidden layers. This layer performs an identity transformation, i.e. it only passes through all the information $y_i=x_i$, where $x_i$ is the output of the previous layers neurons. However, the neurons of the \textit{dropout} layer have a finite probability to for their output to by set to zero, i.e. there is a finite probability that $y_i\coloneqq 0$ for any neuron $i$ in that layer. Since the neurons that are \textit{dropped out} are randomly chosen during each feed-forward training pass, this method not only reduces the overfitting but also introduced stochasticity in the training and often makes the trained model more robust and generalizable.\\
For a much more in-depth explanation on neural networks we refer the interested reader to \cite{Mehta:2018dln}.

\begin{figure}[t]	
 \includegraphics[width=0.48\textwidth]{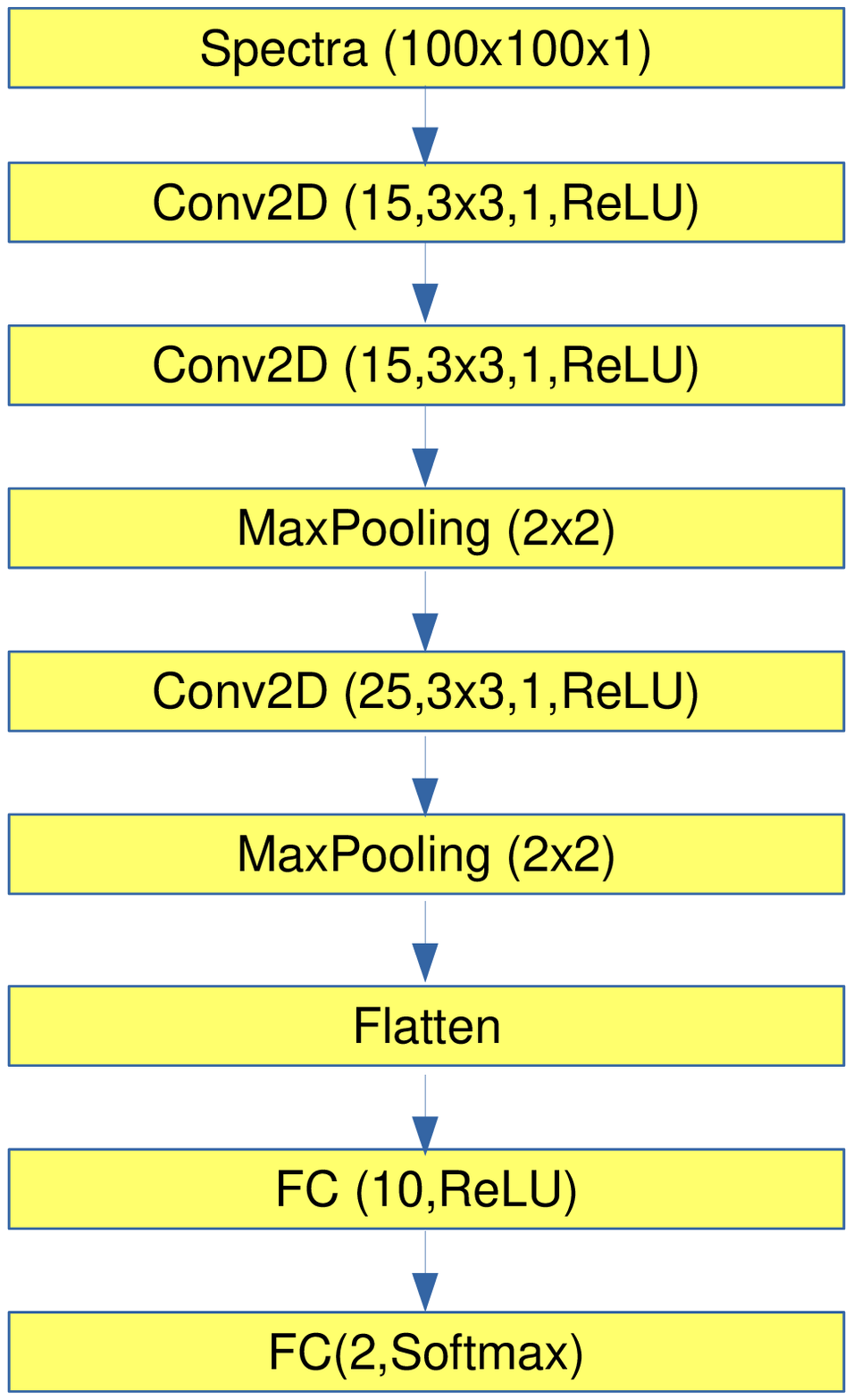}
 \includegraphics[width=0.48\textwidth]{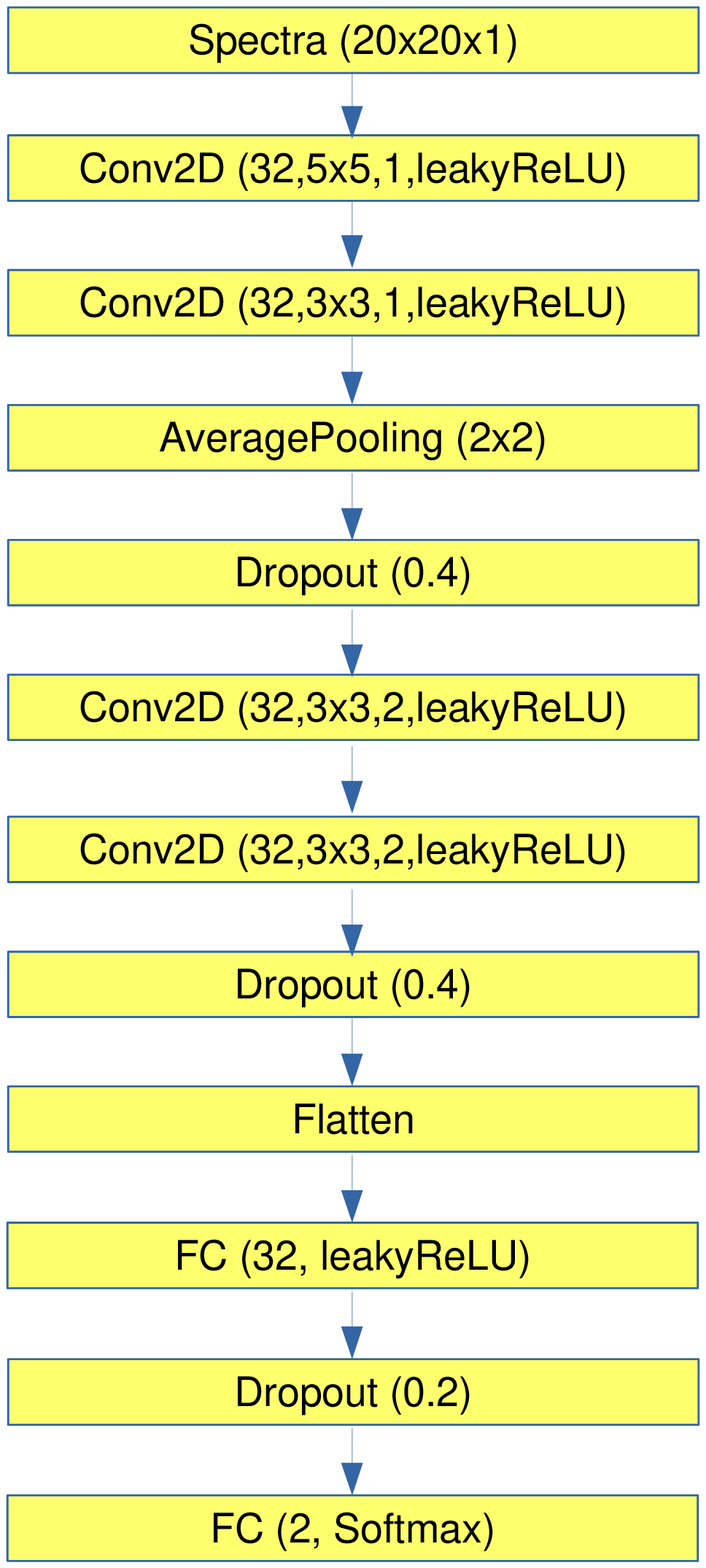}
\caption{\label{cnn1}
Structure of the CNN used in the coordinate space analysis (left) and momentum space analysis (right)).
}
\end{figure}		

\section{Technical details on the Neural Networks used}
In the following, the details on the neural networks and PCA method used are summarized.

\subsection{Convolutional neural network for coordinate space distributions}\label{cnn_coo}
The network employed for the binary classification of the density distributions in the x-y plane, as discussed in section \ref{s_coord}, is structured as shown in figure \ref{cnn1}. It corresponds to a two dimensional convolutional network, using 3 convolutional layers of different size. Between these layers Pooling layers are introduced, which decrease
the dimensionality of the convolutional layers.

The network is constructed and trained using the \textit{Tensorflow} library and \textit{PYTHON}, using the \textit{cross entropy} to calculate the loss function and the \textit{Adam optimizer} for the gradient descent.

\begin{figure*}[t]	
\center \includegraphics[width=0.95\textwidth]{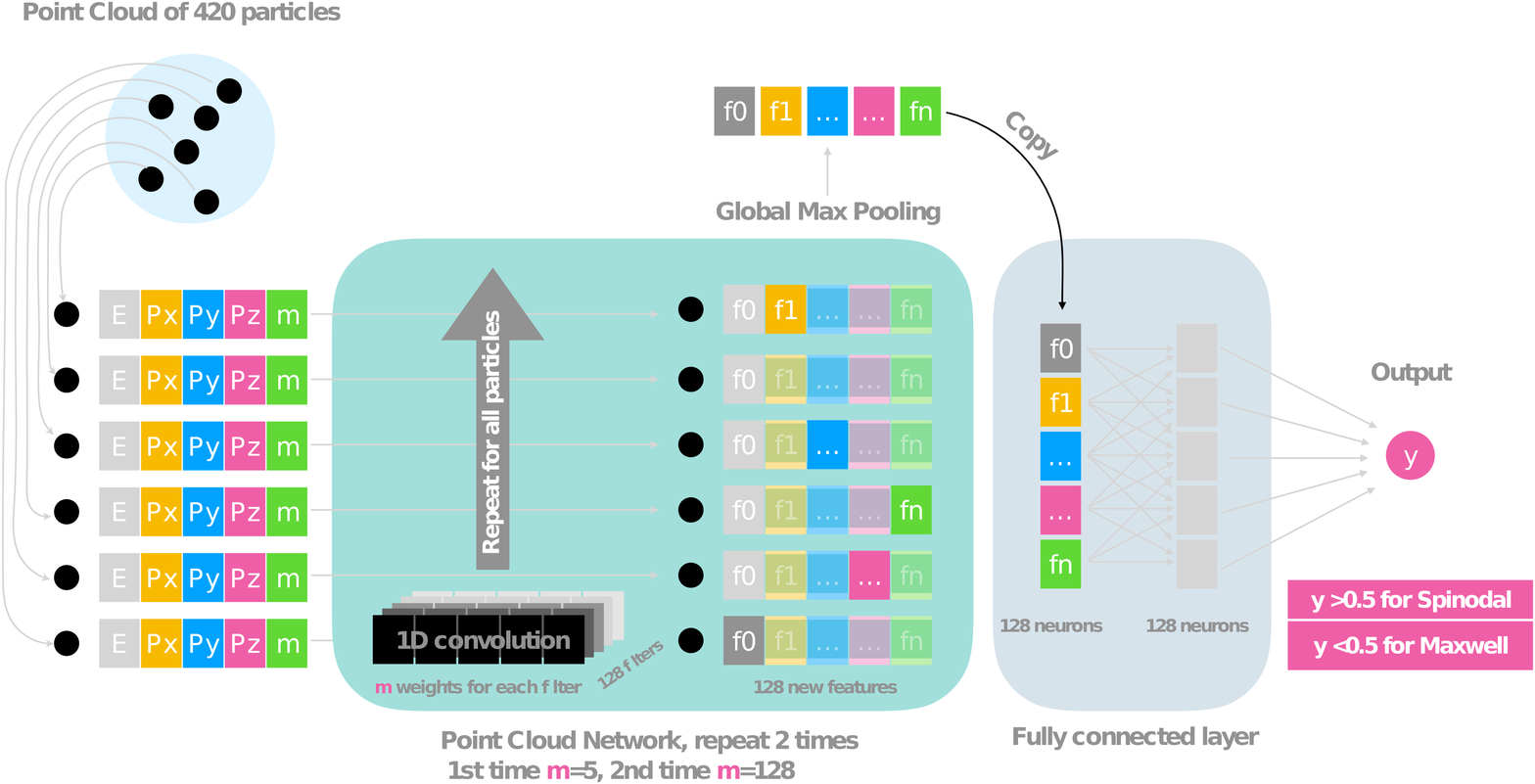}
\caption{\label{pcn_f}
Structure of the Point Cloud Network used in this analysis.
}
\end{figure*}		

\subsection{Convolutional neural network for momentum space distributions}\label{cnn_mom}
The network employed for the binary classification on the momentum density distributions in the $p_x-p_y$ plane, as discussed in section \ref{s_mom}, is structured as shown in figure \ref{cnn1}.

The network is constructed and trained using the \textit{Keras} library and \textit{PYTHON}. Batch normalization and L2 regularization (0.0001) are used. The \textit{Adam optimizer} is used with learning rate 0.00005. These hyperparameters as well as the dropout rates are varied to minimize the overfitting while still allowing the network to capture the important features of the data set.

\subsection{Point-cloud neural network for particle lists}\label{pcn}

This network is employed to do the binary classification using only a list of particle information, i.e. the momentum 4-vector and mass as the coordinate space information cannot be measured directly by the experiments. The input, the network structure and other hyper parameters are listed below and a schematic view is shown in figure \ref{pcn_f}.

\begin{itemize}
    \item For each event, the input is a list of particles. Each list contains 380 to 420 particles, corresponding to the variation of the number of participant baryons in each event, i.e. the length of the input list has to be 420 to allow every event to fit. Each particle has 5 features $(E, p_x, p_y, p_z, {\rm mass})$,
    these features are scaled to approximately lie within the range [0, 1] using the minimum and the maximum values of the first event.
    \item In the point-cloud network, for events with less than 420 baryons, we pad the remaining input array with five-zero vectors.
    \item The \textit{cross entropy} is used to calculate the loss function.
    \item An \textit{SGD optimizer} is used for the gradient descent. For this optimizer, the learning rate is set to 0.0005, the momentum is set to 0.9 and the nesterov is turned on.
     \end{itemize}
The network is constructed and trained using the \textit{Keras} framework and \textit{PYTHON}.\\

\subsection{Fully connected neural network for the momentum difference distribution}
The network employed for the binary classification on the momentum-difference distributions $dN_{pairs}/\Delta p$, as discussed in section \ref{mom_diff_s}, is structured in the following way:

\begin{itemize}
    \item Input are the 200 bin momentum-difference distributions as function of $\Delta p$.
    \item First fully connected hidden layer of 200 neurons and a \textit{leakyReLu} activation function.  
    \item Second fully connected hidden layer of 100 neurons and a \textit{leakyReLu} activation function.  
    \item An output layer with 2 neurons and a softmax activation function.
    \item The learning rate is set to 0.0001 and the \textit{cross entropy} is used as loss function for the training.
     \end{itemize}
     
 The network is constructed and trained using the \textit{Tensorflow} library and \textit{PYTHON}.
 
 \subsection{The principal component analysis (PCA)}\label{pca_intro}
 The PCA used in our analysis is a linear dimensionality reduction using Singular Value Decomposition of the data to project it to a lower dimensional space \cite{minka}, as provided by the \textit{sklearn} library of \textit{PYTHON}. The resulting $x_n$ vectors of the first $n$ components are the principal axes in feature space, representing the directions of the maximum variance in the data.

\end{document}